\documentclass[rmp,aps,twocolumn,superscriptaddress]{revtex4-1}
\usepackage[utf8]{inputenc}
\usepackage{hyperref}
\usepackage{amsmath}
\usepackage{amsfonts}
\usepackage{amssymb}
\usepackage{graphicx}
\usepackage[normalem]{ulem}
\usepackage{siunitx}
\newcommand{\be}{\begin{equation}}
\newcommand{\ee}{\end{equation}}
\newcommand{\bea}{\begin{eqnarray}}
\newcommand{\eea}{\end{eqnarray}}

\newcommand{\bk}{{\bf k}}

\newcommand{\ket}[1]{\left|#1\right>}
\newcommand{\bra}[1]{\left<#1\right|}

\newcommand{\nn}{\nonumber \\ }

\newcommand{\nne}{\nonumber \\ & = &}

\newcommand{\nnp}{\nonumber \\ & & +}

\newcommand{\nnt}{\nonumber \\ & & \times}

\usepackage{color}
\definecolor{mygreen}{rgb}{0,0.5,0} 
\definecolor{mygrey}{rgb}{0.5,0.5,0.5} 
\definecolor{myred}{rgb}{0.75,0,0} 
\definecolor{myblue}{rgb}{0,0,0.75} 
\definecolor{myviolet}{rgb}{0.5,0,0.5} 
\definecolor{mymagenta}{cmyk}{0,1,0,0.12} 
\definecolor{mycyan}{cmyk}{1,0,0,0.12} 
\definecolor{myorange}{rgb}{1,0.5,0}

\newcommand{\Ttot}{T_{\rm acq}}

\newcommand{\br}{{\bf r}}

\usepackage[resetlabels]{multibib}
\newcites{SI,ME}{nada,nada}

\newcommand{\myaffiliation}{\affiliation}
\newcommand{\ICFO}
{
\myaffiliation{ICFO-Institut de Ciencies Fotoniques, The Barcelona Institute of Science and Technology, 08860 Castelldefels, Barcelona, Spain}}
\newcommand{\ICREA}
{
\myaffiliation{ICREA -- Instituci\'o Catalana de Recerca i Estudis
Avan\c{c}ats, 08010 Barcelona, Spain}
}

\begin{document}
 
\newcommand{\mytitle}{On the $\hbar$ energy resolution limit in field sensing } 
\renewcommand{\mytitle}{Quantum limits to the energy resolution of magnetic field sensors } 
\title{\mytitle
 }
\author{Morgan W. Mitchell}
\ICFO
\ICREA
\author{Silvana Palacios Alvarez}
\ICFO

\date{\today}

\begin{abstract}
The energy resolution per bandwidth $E_R$ is a figure of merit that combines the field resolution, bandwidth or duration of the measurement, and size of the sensed region. Several  different dc magnetometer technologies approach $E_R = \hbar$, while to date none has surpassed this level. This suggests a technology-spanning quantum limit, a suggestion that is strengthened by model-based calculations for nitrogen-vacancy centres in diamond, for superconducting quantum interference device (SQUID) sensors, and for some optically-pumped alkali-vapor magnetometers, all of which predict a quantum limit close to $E_R = \hbar$.  Here we review what is known about energy resolution limits, with the aim to understand when and how $E_R$ is limited by quantum effects. We include a survey of reported sensitivity versus size of the sensed region for  more than twenty magnetometer technologies, review the known model-based quantum limits, and critically assess possible sources for a technology-spanning limit, including zero-point fluctuations, magnetic self-interaction, and quantum speed limits.  Finally, we describe sensing approaches that appear to be unconstrained by any of the known limits, and thus are candidates to surpass $E_R = \hbar$.  
\end{abstract}
\maketitle

\tableofcontents

\newcommand{\BW}{\Delta \nu}
\newcommand{\Btrue}{B_{\rm true}}
\newcommand{\ERL}{ERL}
\newcommand{\hERL}{$\hbar$~\ERL}
\newcommand{\szi}{\tilde{s}_{z,i}}
\newcommand{\szj}{\tilde{s}_{z,j}}
\newcommand{\supz}{^{(z)}}
\newcommand{\supjk}{^{(jk)}}
\newcommand{\supik}{^{(ik)}}
\newcommand{\supk}{^{(k)}}

\newcommand{\rhat}{{\bf  r}}
\newcommand{\bdir}{{\bf b}}
\renewcommand{\rhat}{\pmb{\scriptr}}
\renewcommand{\rhat}{{\mathbb R}}
\newcommand{\zhat}{{\mathbb Z}}
\newcommand{\Slim}{S_{\rm lim}}
\newcommand{\by}{{\bf y}}
\newcommand{\omdd}{\omega_{\rm dd}}
\newcommand{\omL}{\omega_{L}}

\newcommand{\limsym}{{\cal S}}
\newcommand{\name}{{noise power spectral density}}
\newcommand{\acr}{{NPSD}}
\newcommand{\ArticleType}{article~}
\renewcommand{\ArticleType}{colloquium~}

\section{Introduction}
Low-frequency magnetic fields are ubiquitous and provide important information in applications ranging from nanotechnology \cite{AriyaratneNC2018} to brain studies \cite{BotoNI2017} to space science \cite{ArridgeNP2016}. A plethora of sensing scenarios, and the availability of many physical systems with strong magnetic response, have led to many distinct magnetometer technologies. A recent and extensive review can be found in \cite{GroszBook2016}. It is of both fundamental and practical interest to know how well quantum physics allows such sensors to perform.  At a fundamental level, prior work on quantum limits of sensing has uncovered connections to the geometry of quantum states \cite{BraunsteinPRL1994}, entanglement in many-body systems \cite{SorensenPRL2001}, quantum information processing \cite{GiovannettiPRL2006, RoyPRL2008}, and quantum non-locality \cite{TuraS2014,SchmiedS2016}. These results for the most part concern quantum estimation theory \cite{HelstromJSP1969, Helstrom1976} applied to generalized linear interferometers \cite{LeeJMO2002}. High-performance magnetometers, however, employ methods not easily mapped onto linear interferometry \cite{MitchellQST2017}, and one may hope that understanding their quantum limits will yield still other fundamental insights.  

In this \ArticleType we focus on energy resolution limits (ERLs), which are constraints on the energy resolution per bandwidth $E_R$, a figure of merit that combines field resolution, measurement duration or bandwidth, and the size of the sensed region \cite{RobbesSAA2006}.  $E_R$ has units of action, with smaller values indicating a better combination of speed, size and sensitivity.  The best magnetometer technologies now reach $E_R \approx \hbar$. ERLs near this value are predicted for important magnetometry technologies, including superconducting quantum interference devices (SQUIDs) \cite{KochPRL1980},  optically-pumped magnetometers (OPMs) \cite{JimenezBook2017} and spin-precession sensors with fixed, random spin positions, e.g. nitrogen-vacancy centers in diamond (NVD) \cite{MitchellERLARX2019}.  The nature and scope of ERLs is thus also a practical question that informs efforts to improve energy resolution beyond the current state of the art. 

One of the most intriguing features of ERLs is the suggestion that there may be a single, technology-spanning ERL, one that constrains any magnetic field measurement, regardless of how it is performed. This suggestion emerges most immediately from a multi-technology survey of reported sensitivities (see Figure~\ref{fig:SlowMagSensitivity} and Section~\ref{sec:LiteratureValues}), all of which obey $E_R \ge \hbar$  even as some come  close to this level.  Given that the known limits for SQUIDs, OPMs and NVD sensors are also near this level, it is natural to ask whether these ERLs could be distinct manifestations of a single, technology-spanning ERL.  Such a limit could  plausibly be imposed by general quantum limits, for example the Margolus-Levitin bound \cite{MargolusPFWPC1998}, which relates the speed of evolution to the available energy, or the Bremermann-Bekenstein bound \cite{BekensteinPRD1981, BremermannIJTP1982}, which relates the entropy and thus information content of a region to its energy content and size.  

The objective of this \ArticleType is to bring together, and when possible to synthesize, the many dispersed insights  that bear on the question of when and how quantum mechanics constrains the energy resolution of a magnetic field sensor. The text is organized as follows: In Section \ref{sec:History} we describe ERLs as they appear in the scientific literature for different sensor types. In Section \ref{sec:NatureAndForm}, we discuss the physical meaning of an ERL and note its relation to independence of quantum noise sources. Section \ref{sec:LiteratureValues} provides a survey of reported sensitivities.  In Section \ref{sec:KnownLimits} we present model-based ERLs for SQUIDs, alkali vapor OPMs, and color center (e.g. NVD) sensors.   In Section \ref{sec:TechnologyIndependent} we assess technology-independent quantum limits, e.g. quantum speed limits, and their potential to supply a technology-spanning ERL.  In Section \ref{sec:ProposedToBeat} we describe sensing approaches that evade the known limits, and thus may have potential to surpass current state-of-the-art energy resolutions. 

\begin{figure}[t]
 \includegraphics[width=0.6 \columnwidth]{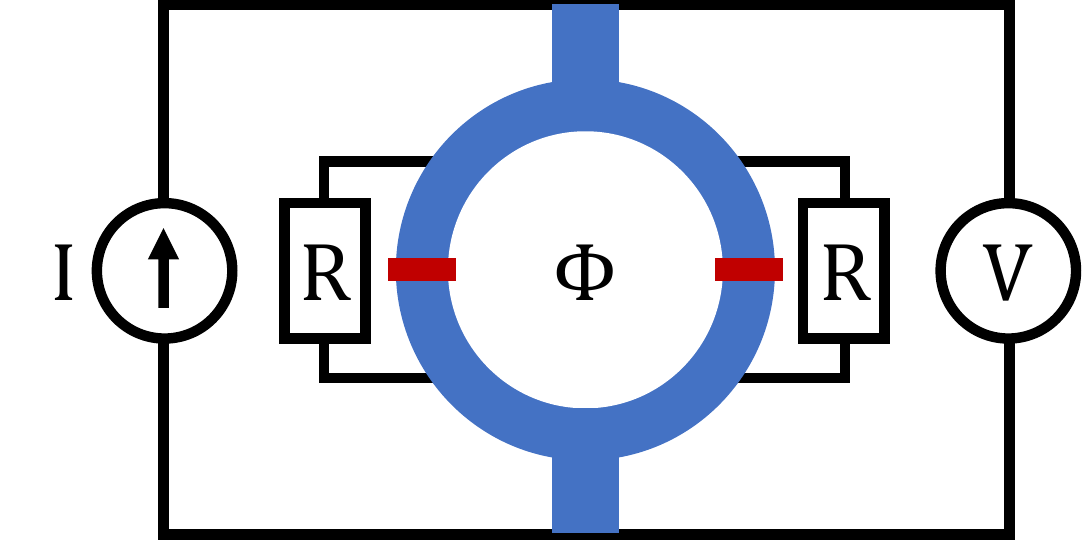}
\caption{Schematic diagram of a dc SQUID in constant current mode with resistively shunted Josephson junctions.  A superconducting loop (blue) is interrupted by two Josephson junctions (red). The junctions are shunted by resistances $R$.  A constant current $I$ feeds the SQUID, and the generated voltage $V$ is used to infer the flux $\Phi$ threading the SQUID loop.  
 }
 \label{fig:SQUID}
\end{figure}

\section{History and origins}
\label{sec:History}

The ERL concept emerged from an analysis of dc SQUID sensitivity by Tesche and Clark (TC) \cite{TescheJLTP1977, KochPRL1980}. A SQUID is a planar field sensor based on the Josephson effect and measures the flux $\Phi$ through a loop of superconducting material, see Fig.~\ref{fig:SQUID}. A  more detailed description is given in Section \ref{sec:KnownLimits}. TC  considered a lumped-circuit model of the dc SQUID dynamics, to compute the equivalent flux-noise power spectral density ${S_\Phi(\nu)}$ of $\Phi$, where $\nu$ is the linear frequency. Optimization of the SQUID parameters for best dc sensitivity, i.e. for minimum $S_\Phi(0)$, yeilds the bound 
\begin{equation}
\label{eq:ERLForSQUIDs}
E_R^{(\rm dc~SQUID)} \equiv \frac{S_\Phi(0)}{2 L} \ge \hbar,
\end{equation}
where $L$ is the inductance of the SQUID pickup loop\footnote{The 1977 TC publication assumed quantum noise arising from electron pair shot noise in the current across the Josephson junctions, and arrived to a limit of $h/2 = \pi \hbar$. Later analyses assume quantum noise from zero-point fluctuations in resistances shunting the junctions, and arrive to a limit of $\hbar$ \cite{KochAPL1981,RobbesSAA2006}.}.  The name ``energy resolution'' was applied to $E_R = S_\Phi(0)/(2L)$  by analogy to $\Phi^2/(2L)$, the magnetostatic energy in a current loop.  As should be clear from Eq.~(\ref{eq:ERLForSQUIDs}), $E_R$ has units of action, not energy, and in more recent literature $E_R$ is  referred to as the ``energy resolution per bandwidth.''  

To compare against other kinds of sensors, it is interesting to have a purely geometric expression for this limit.  To this end, we note that 
$\Phi = B A$, where $A$ is the loop area, and that $L = \sqrt{A} \mu_0/\alpha$, where $\alpha$ is a wire geometry factor of order unity.  We can thus re-express the TC limit as 
\begin{equation}
\label{eq:ERLForArea}
E_R^{(\rm area)} \equiv \frac{S_B(0) A^{3/2}}{2 \mu_0} \ge \alpha \hbar.
\end{equation}

The use of energy resolution as a measure of sensitivity has spread to other areas, including both BEC and hot-vapor OPMs \cite{VengalattorePRL2007, Dang2010, GriffithOE2010, JimenezBook2017}, and cross-technology reviews \cite{BendingAP1999, RobbesSAA2006, YangPRAp2017}.  For a planar BEC sensor, the geometrical form of the ERL, i.e. Eq.~(\ref{eq:ERLForArea}), has been directly used for an inter-technology comparison \cite{VengalattorePRL2007}.  For volumetric sensors, the energy resolution has been defined with reference to $B^2V/(2\mu_0)$, the magnetostatic energy in a volume $V$, to give
\begin{equation}
\label{eq:ERLvolCont}
E_R^{(\rm vol)} \equiv \frac{S_B(0)V}{2 \mu_0}.
\end{equation}
  A first-principles study of a possible quantum bound on $E_R^{(\rm vol)}$ for OPMs, and comparison against the planar ERL for SQUIDS, appears to have been reported by Lee and Romalis \cite{LeeDAMOP2008}. See also Romalis \textit{et al.} \cite{RomalisPRLc2014}.  

\newcommand{\Nrep}{N_{\rm rep}}
\newcommand{\leff}{l_{\rm eff}}

\section{Nature and form of energy resolution limits}
\label{sec:NatureAndForm}

To understand the meaning of an ERL, it is convenient to consider, rather than a continuous measurement, a sequence of discrete field measurements, averaged to obtain the dc field value.  Consider a magnetic sensor of volume $V$ that, after observation time $T$, gives a reading $B_{\rm obs} = B_{\rm true} + \delta B$, where $B_{\rm true}$ is the true value of the field and $\delta B$ is the measurement error.  We assume that through calibration of the sensor $\langle \delta B \rangle = 0$, such that $B_{\rm obs}$ is an unbiased estimator for $B_{\rm true}$. 
The mean apparent magnetostatic energy in the sensor volume is 
\begin{eqnarray}
\label{eq:EObs}
E_{\rm obs} &=& \langle B^2_{\rm obs}\rangle V/(2 \mu_0) 
\nne  B_{\rm true}^2 V/(2 \mu_0) +  \langle \delta B^2\rangle V/(2 \mu_0).  
\end{eqnarray}

We now consider performing such $T$-duration measurements as often as possible, i.e. with measurement repetition period $T$, and averaging them. We assume the measurement is subject to only quantum noise, all other noise sources having been reduced to negligible levels. We can then use the statistical independence of quantum noise in two ways. First, we note that $\langle \delta B^2 \rangle T = S_B(0)$, such that the second term in Eq.~(\ref{eq:EObs}) becomes  $E_R^{({\rm vol})}$.  We thus recognize  $E_R^{({\rm vol})}$ as the bias in the magnetostatic energy estimate. 

Second, we note that 
 in total acquisition time $\Ttot$ we can average $\Nrep \equiv \Ttot/T$ measurements, to yield a variance $\langle \delta B^2 \rangle/\Nrep =\langle \delta B^2 \rangle T/\Ttot$. Given a fixed $\Ttot$ and a choice among measurements with different values for $\langle \delta B^2 \rangle$ and $T$, the measurement with the smallest $\langle \delta B^2 \rangle T$ 
is thus the superior measurement.  If a quantum limit exists, it is because there is a limit on the product $\langle \delta B^2 \rangle T$.
 The same argument applies to the volume, if we imagine filling a volume with non-overlapping sensors and averaging their readings.  As a result, a volumetric ERL, if it is of quantum origin, must take the form
\begin{equation}
\label{eq:ERLvolDiscr}
E_R^{({\rm vol})}  \equiv \frac{S_B(0)V}{2 \mu_0} =  \frac{\langle \delta B^2\rangle VT } {2 \mu_0}  \ge \limsym, 
\end{equation}
where  $\limsym$ is a constant with units of action.  We use the common shorthand $\delta B$ for $\langle \delta B^2 \rangle^{1/2}$ and refer to $\delta B \sqrt{T}$ as the sensitivity.

\begin{figure*}[t]
 \includegraphics[width=0.95 \textwidth]{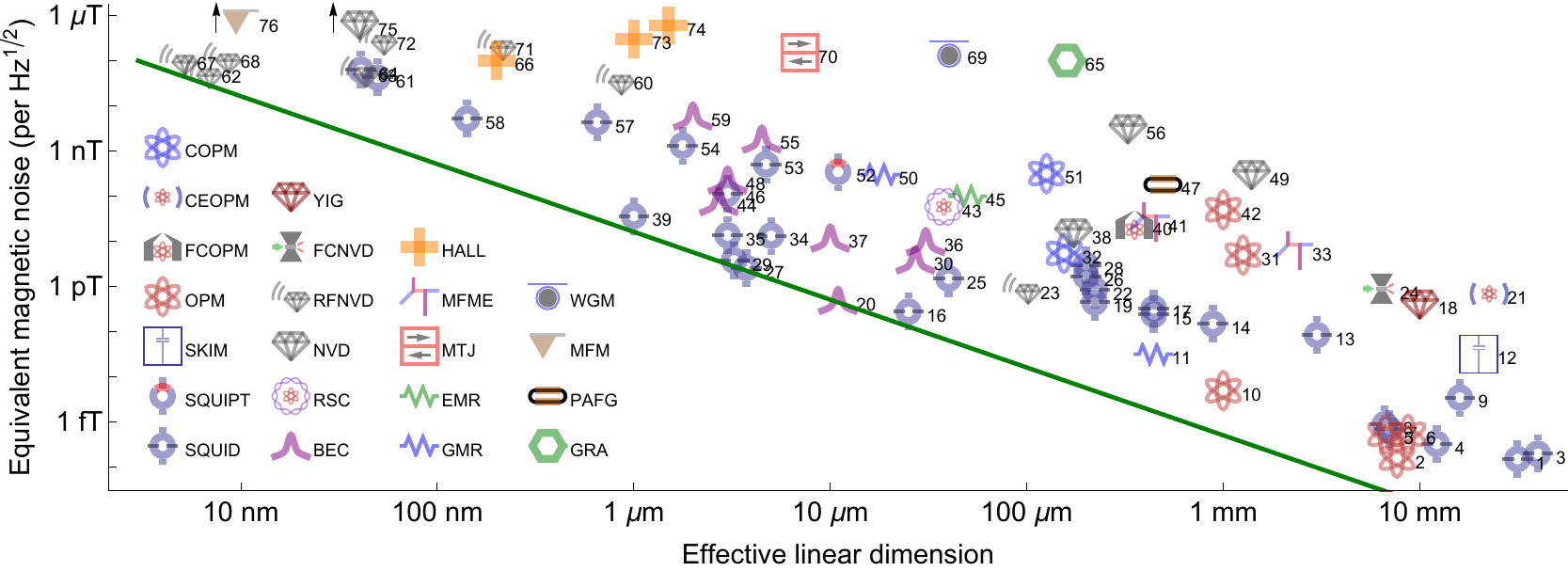}
\caption{Reported magnetic sensitivity $\delta B \sqrt{T}$ for different sensor technologies versus size of the sensitive region.  Effective linear dimension $\leff$ indicates $\sqrt{{\rm area}}$  
 for planar sensors and $\sqrt[3]{{\rm volume}}$  
 for volumetric ones. For point-like systems such as single spins, $\leff$ indicates $\sqrt[3]{\rm volume}$ for a sphere with radius equal to the
 minimum source-detector distance.  For works reporting sensitivity in units of magnetic dipole moment, we convert to field using the reported sample distance. Excepting RFNVD, noise levels are the lowest reported value at frequency $\le \SI{1}{\kilo\hertz}$.  Arrow ($\uparrow$) indicates value is off scale.  SQUID - superconducting quantum interference device; SQUIPT - superconducting quantum interference proximity transistor; SKIM - superconducting kinetic impedance magnetometer;  OPM - optically-pumped magnetometer;  FCOPM - OPM with flux concentrators;  CEOPM - cavity-enhanced OPM;  COPM - OPM with cold thermal atoms;  BEC Bose-Einstein condensate; RSC - Rydberg Schr\"{o}dinger cat; NVD - nitrogen-vacancy center in diamond; RFNVD radio-frequency NVD; FCNVD - NVD with flux concentrators;  YIG yttrium-aluminum-garnet;  GMR - giant magneto-resistance;  EMR extraordinary magneto-resistance; MTJ - magnetic tunnel junction;   MEMF - magnetoelectric multiferroic;  HALL - Hall effect sensor;  GRA - graphene; PAFG - parallel gating fluxgate; MFM - magnetic force microscope,  WGM - whispering-gallery mode magnetostrictive.  Line shows $E_R \equiv \langle \delta B^2 \rangle T \leff^3/(2\mu_0) = \hbar$.  
 Numeric labels refer to Table \ref{tab:SensSizeTable}.   
 }
 \label{fig:SlowMagSensitivity}
\end{figure*}

\section{Scaling of sensitivity with extent of the sensed region}
\label{sec:LiteratureValues}

High-sensitivity magnetometers have been demonstrated or proposed with all possible dimensionalities:  point-like, linear, planar, and volumetric.  Examples of point-like sensors are single NVDs \cite{FangPRL2013, TrusheimNL2014, LovchinskyS2016, AriyaratneNC2018} and single trapped ions \cite{RusterPRX2017}.  Linear sensors include ferromagnetic needles \cite{JacksonKimballPRL2016, BandPRL2018} and some cold atomic ensembles \cite{Sewell2012, Behbood2013}.  Planar sensors include superconducting sensors of various types  \cite{RobbesSAA2006, GiazottoNP2010, KherEUCAS2013, LuomahaaraNC2014, KherJLTP2016, KherThesis2017}, Hall-effect sensors \cite{BendingAP1999} and several others \cite{RobbesSAA2006, GroszBook2016}.  Volumetric sensors include OPMs \cite{KominisN2003, Dang2010,GriffithOE2010,SavukovBook2017,WeisBook2017,JimenezBook2017,GawlikBook2017}, ensemble NVD sensors \cite{WolfPRX2015, BarryPNAS2016, JensenBook2017}, and others.  Sensors employing trapped Bose-Einstein condensates \cite{WildermuthAPL2006, VengalattorePRL2007, YangPRAp2017}  or cold thermal ensembles may approximate any of these geometries, depending on the trap configuration.

\newcommand{\lmin}{l_{\rm min}}

Regardless of the sensor dimensionality, the field to be detected exists in three-dimensional space, and moreover varies smoothly in that space, except near magnetic sources.  Because of this, a sensor's reading is representative of the field in a three-dimensional volume, even if the sensor is of lower dimensionality.  For example, we may consider a point-like sensor embedded in a support that prevents magnetic sources to approach closer than a minimum distance $\lmin$.  By $\nabla \cdot {\bf B} = 0$, the field experienced by the sensor is equal to the average of the field inside a sphere of radius $\lmin$ about the sensor position. The sensor signal is the same as it would be, were the sensor to uniformly sample  this spherical volume.  We can thus assign an effective volume $4\pi \lmin^3/3$ to the point-like sensor.
   To enable a uniform comparison of different sensor types, we define $\leff \equiv \sqrt[3]{\rm volume}$ and $\leff = \sqrt{\rm area}$, respectively, for the effective linear dimensions of volumetric and planar sensors.  Noting that with these definitions Eq.~(\ref{eq:ERLForArea}) and Eq.~(\ref{eq:ERLvolCont}) coincide, we can hypothesize the technology-spanning ERL 
\begin{equation}
\label{eq:ERLGeneral}
E_R  \equiv \frac{S_B(0)\leff^3}{2 \mu_0} =  \frac{\langle \delta B^2\rangle \leff^3 T } {2 \mu_0}  \ge \alpha \hbar,
\end{equation}
where $\alpha$ is again a number of order unity.

In Fig.~\ref{fig:SlowMagSensitivity} we show sensitivity versus effective linear dimension for many representative publications on high-sensitivity magnetic field detection.  Only measured sensitivities are included, and only when the dimensions of the sensitive region could be determined. With a few exceptions (to be remarked below), the survey is restricted to dc field sensors, which we take to include field components below \SI{1}{\kilo\hertz}.  While most sensors operate continuously, a number of sensors in the survey operate in a pulsed mode. For example, cold atom experiments take time to accumulate atoms prior to any sensing, making their cycle time longer than the measurement time. Because we are concerned here with fundamental limits, when computing the energy resolution per bandwidth we include such delays only if they appear unavoidable for fundamental reasons. For example, atom-trap loading time is not included, because one can imagine ways to deliver a new batch of atoms each time the previous batch is consumed.  In contrast, delays associated with optical pumping or fluorescence detection (both of which require time for spontaneous emission) would appear unavoidable. 

For works that report a field-equivalent noise spectral density $S_B(\nu)$ and the dimensions of the sensitive region, no conversion of physical quantities is required.  For works that report sensitivity in units of magnetic moment $\mu$, e.g. for magnetic microscopy applications, we convert the equivalent noise $S_\mu(\nu)$ to field units using the sample-sensor distance assuming a dipole field distribution.

As seen from Fig.~\ref{fig:SlowMagSensitivity}, several  different technologies come close to $E_R = \hbar$. These include micro-SQUIDs \cite{Cromar1981, VanHarlingenAPL1982, AwschalomAPL1988, WakaiAPL1988, MuckAPL2001}, spinor Bose-Einstein condensates \cite{VengalattorePRL2007}, and SERF-regime OPMs \cite{DangAPL2010, GriffithOE2010}.  As rf magnetometers, single NV centers in diamond \cite{LovchinskyS2016} 
also are close to $E_R = \hbar$.

\section{Known limits for specific technologies}
\label{sec:KnownLimits}

For three  well-studied high sensitivity magnetometer types, model-based calculations are known to lead to an ERL.  We have already mentioned the TC limit for dc SQUID sensors, the origin of Eq.~(\ref{eq:ERLForSQUIDs}).  More recent calculations for spin-precession sensors, both OPMs and NVD sensors, give rise to a limit on the volumetric energy resolution of Eq.~(\ref{eq:ERLvolCont}).  Although derived for different systems using different models, these limits appear to agree.

\subsection{dc SQUID sensors}
\label{sec:TCLimit}

The dc SQUID analyzed by TC \cite{TescheJLTP1977} consists of a loop of superconducting material interrupted by two Josephson junctions (JJs), as illustrated in Fig.~\ref{fig:SQUID}, with a constant current bias $I$ and resulting voltage $V$ across the SQUID.  The dynamics of the dc SQUID are in general complex and nonlinear, but in some regimes the SQUID provides a direct relationship between the flux $\Phi$ threading the loop and $V$, allowing $\Phi$ to be inferred from $V$, which can be measured using low-noise amplifiers.  In other regimes, the flux-current relationship is hysteretic and the flux cannot be simply inferred from $V$. To avoid these hysteretic regimes, damping is typically introduced in the form of resistances shunting the JJs.  These shunt resistances introduce both thermal noise (Johnson-Nyquist noise) and quantum noise (zero-point current fluctuations) into the SQUID dynamics.  TC's analysis showed that if the resistances provide enough damping to make the flux-current relation single valued, then they also introduce enough quantum noise to impose an energy resolution limit, as in Eq.~(\ref{eq:ERLForSQUIDs}). The TC analysis has been extended to more detailed dc SQUID models \cite{KochPRL1980, WakaiAPL1988, RyhanenJLTP1989}, and is reviewed by Robbes \cite{RobbesSAA2006}.   With careful construction, small dc SQUID devices have reported ${S_\Phi(0)}/({2 L})$ as low as $2 \hbar$ \cite{AwschalomAPL1988, WakaiAPL1988, MuckAPL2001}. 

\newcommand{\psinaught}{| \psi(0) \rangle}
\newcommand{\unkn}{{\cal B}}
\newcommand{\Topt}{T_{\rm opt}}
\newcommand{\subSD}{_{\rm SD}}

\subsection{Alkali-vapor optically-pumped magnetometers}
\label{sec:OPMLimit}
A  different magnetic sensing technology, the hot vapor OPM \cite{BudkerNP2007, JimenezBook2017, WeisBook2017, JensenBook2017, SavukovBook2017} has been shown to obey a volumetric energy resolution limit.  In these devices, quantum noise in the form of optical shot noise, optical spin-rotation noise, and spin projection noise all contribute to the effective magnetic noise \cite{SmullinPRA2009} and scale differently with atomic number density, volume, and optical probe power.  When optimized for sensitivity, and in the most sensitive, spin-exchange-relaxation-free regime, the equivalent magnetic noise is limited by 
\begin{equation}
\label{eq:SD}
S_B(0) \ge \frac{1}{\gamma^2} \frac{\bar{v} \sigma\subSD }{V },
\end{equation}
where $\gamma$ is the gyromagnetic ratio, $\bar{v}$ is the thermal velocity, $\sigma\subSD$ is the cross-section for spin-destruction collisions and $V$ is the volume of the sensor \cite{JimenezBook2017}.  When expressed in terms of the field energy, this quantum limit is
\begin{equation}
\label{eq:ERLForOPMs}
\frac{S_B(0) V}{2 \mu_0} \ge \frac{\bar{v} \sigma\subSD}{2 \mu_0 \gamma^2}.
\end{equation}

Spin-destruction collisions result when spin angular momentum is transferred to the centre-of-mass degree of freedom. In $^{87}$Rb, the limiting energy resolution is  nearly $\hbar$ when calculated using measured spin-destruction rates \cite{JimenezBook2017}.  This agreement with the $E_R$ of dc SQUIDs appears to be a coincidence, as other alkali species and noble gases have significantly lower values for $\bar{v}\sigma_{\rm SD} \gamma^{-2}$, see Section \ref{sec:SDRates}. 
Realized OPMs using $^{87}$Rb and K have demonstrated $E_R$ as low as {44}~{$\hbar$} \cite{AllredPRL2002, SavukovPRL2005, ShahNPhot2007, LedbetterPRA2008, SmullinPRA2009, DangAPL2010}, limited by environmental noise.

\subsection{Immobilized spin ensembles, e.g. nitrogen-vacancy centers in diamond}
\label{sec:NVDLimit}
\label{sec:NVs} In contrast to OPMs, which employ mobile spins in the vapour phase, sensors employing nitrogen-vacancy (NV) centres in diamond use immobile spins fixed in a solid matrix  \cite{DohertyPR2013, AcostaBookChapter2013, RondinRPP2014}.  In this scenario, no entropy is input to the spin system by collisions, and decoherence due to nuclear spins \cite{TaylorNP2008} can in principle be fully eliminated by use of isotopically-pure $^{12}$C diamond \cite{BalasubramanianNMat2009}. Nonetheless, the spins necessarily interact with each other by dipole-dipole coupling, and this interaction allows angular momentum loss to the crystal lattice.  A recent analysis of the limiting sensitivity imposed by this effect \cite{MitchellERLARX2019} finds that for spatially-disordered spins, dipolar coupling of the sensor spins themselves is sufficient to cause depolarization and enforce Eq.~(\ref{eq:ERLGeneral}) with $\alpha \approx 1/2$.  

\section{Possible sources of technology-independent limits}
\label{sec:TechnologyIndependent}

The fact that three  different sensor technologies all arrive to the same limit again suggests there could be a more general, technology-spanning reason for energy resolution to be limited to $\hbar$.  As examples, the magnetic field itself is subject to quantum fluctuations, and any system measuring the field must obey quantum speed limits.  In this section we evaluate several general considerations that could give rise to a technology-spanning ERL.

\newcommand{\rSens}{{r_{\rm S}}}
\newcommand{\bff}{{\bf f}}
\newcommand{\MeasDur}{T_{\rm eff}}

\subsection{Standard quantum limit, Heisenberg limit and amplification quantum noise}
\label{sec:QuantumMetrology}

A considerable literature has developed around the topic of ``quantum-enhanced sensing,'' \cite{GiovannettiS2004, PezzeRMP2018,  BraunRMP2018} which explores the use of non-classical states to achieve sensitivity between the standard quantum limit (SQL) and the so-called Heisenberg limit (HL).  These limits are defined in terms of the number of particles employed in the sensing procedure.  For example, the SQL for measurement of a phase angle $\phi$ with $N$ non-interacting two-level systems is $\langle \delta \phi^2 \rangle = N^{-1}$.  It is notable that the ERL makes no reference to particle number, whereas the SQL and HL make no reference to spatial extent or to time.  Any derivation of the ERL from the SQL or HL would thus require an important input from other physical principles.  

Similarly, linear, phase-insensitive amplification is known to introduce quantum noise intrinsic to the amplification process \cite{CavesPRD1982}. This noise, when present, has a magnitude similar to the intrinsic quantum noise of the system to be measured, resulting in an equivalent noise comparable in magnitude to the SQL. As such, amplification quantum noise is not \textit{per se} an explanation of the ERL.

\subsection{Energy-time uncertainty relation} 
\label{sec:DeltaEDeltaT}
The energy-time uncertainty relation $\delta E \delta t \ge \hbar/2$ is the paradigmatic example of an uncertainty relation not derived using the commutation relation of the involved quantities. The same form of the energy-time uncertainty relation is found in a variety of scenarios with different meanings for $t$, for example in the question of arrival-time distributions \cite{AharonovPR1961} or in the estimation of a Hamiltonian in limited time \cite{AharanovPRA2002}.  This latter problem is closely related to quantum estimation methods \cite{HelstromJSP1969, GiovannettiS2004}, in which $\delta t$ is the measurement duration, i.e. $T$.  Using this definition and taking $E$ to be the magnetostatic energy $V B^2/(2\mu_0)$, we find the relation 
\be
\label{eq:DeltaEDeltaTLimit}
\frac{\delta(B^2) V T}{2\mu_0} \ge \frac{1}{2} \hbar
\ee
which differs from Eq.~(\ref{eq:ERLvolDiscr}) in that it contains the r.m.s. deviation of $B^2$, rather than the variance of $B$.  This is an important difference: Considering a small uncertainty $\delta B$ on a large field $B \gg \delta B$, $\delta (B^2) \approx 2 B \delta B$, so that Eq.~(\ref{eq:DeltaEDeltaTLimit}) allows for arbitrarily small $\delta B$ as $B \rightarrow \infty$.  It would appear that the energy-time uncertainty relation places no limit on magnetic sensitivity.

\subsection{Zero-point and thermal field fluctuations} 
\label{sec:ZeroPoint}
To our knowledge, not much attention has been paid to the possibility that the quantum fluctuations of the magnetic field might be detectable and thus a source of noise for sensitive magnetometry.  One can imagine a scenario in which the goal is precisely to observe the zero-point fluctuations, in which case these fluctuations are not noise, but rather signal.  Such a measurement has indeed been reported with THz electric fields \cite{TighineanuPRL2014, RiekS2015}.  For our present purposes, however, we are more interested in the effect of zero-point fluctuations when measuring fields of material origin, e.g. from a current.  In such a measurement the zero-point fluctuations would be considered noise, if indeed they contribute to the recorded signal.  

In Appendix~\ref{app:ZeroPoint} we analyze the following model for the effect of zero-point fluctuations.  First, we define a spherical region ${\cal R}$ of radius $\rSens$ and volume $V$, and  the ${\cal R}$-averaged field component $\bar{B}_z(t) \equiv V^{-1} \int d^3r \, \rho({\br}) B_z(\br,t)$, where $\rho({\br})$ is a weighting function. At any value of the time $t$, $\bar{B}_z(t)$ is a hermitian operator and thus a valid quantum mechanical observable.  An ideal measurement of this observable will have a mean that is the true spatially-averaged value of any externally-applied field in that region, and finite contribution to the variance from zero-point fluctuations.  Measurement of this observable will also produce a random and in principle unbounded measurement back-action in the conjugate electric field ${\bf E}$, which through the Maxwell equations will propagate into ${\bf B}$.  Nonetheless, due to the space-bounded nature of the measurement, and the fact that EM fields propagate at speed $c$, all effects of this disturbance will propagate out of the volume of interest after a time $\MeasDur = 2\rSens/c$, at which point a measurement of $\bar{B}(t + \MeasDur)$, with independent noise, can be made.  Considering a sequence of such measurements separated by $\MeasDur$, we find a zero-point-limited energy resolution 
\begin{eqnarray}
 E_R & \ge & \alpha  \hbar,
\end{eqnarray}
where $\alpha \approx 1.3$ for the specific weighting considered (see \autoref{eq:WeightingFunction}). This is not far from other derived energy resolution limits.  Within the same analysis we consider thermal fields and find a contribution to $E_R$ of  $\approx   \rSens  k_B T_{B}/c$, where $T_{B}$ is the field temperature. 

However elegant such a solution to the energy resolution question might appear, we believe it hides a subtle and incorrect assumption, and does not represent a fundamental limit.  If we replace the ``ideal measurement'' of $\bar{B}$ with a more detailed model of the measurement, the problem becomes apparent.   Imagine we fill the region ${\cal R}$ with a rigid sphere of zero temperature, uniformly magnetized material (e.g. a single ferromagnetic domain). We assume the energy of the ferromagnet is independent of its spin direction, and that it is ``stiff'' in the sense that all regions must share a single spin orientation. The measurement consists of allowing this object to freely  rotate in response to the field it experiences, and then observing its new orientation.  If the initial state of this evolution is a product of the ground state of the magnet and the ground state of the field (vacuum), we would indeed observe a random  rotation, a noise signal due to the zero-point magnetic field. If, on the other hand, we take as our initial condition the minimum-energy state of the coupled field-ferromagnet system, we would -- simply because we are considering an energy eigenstate --  observe no  rotation, nor indeed change of any kind.  From such an initial condition, any rotation of the system, even if stochastic, would at least temporarily violate angular momentum conservation. See Appendix \ref{sec:QuantumSpinVacuum} for a more detailed discussion of this point.  We conclude that this in-principle-possible magnetic sensor, embedded in vacuum and allowed to find equilibrium with it, would not experience a noise from vacuum fluctuations.

\subsection{Spin intrinsic-noise self-interaction }
\label{sec:ProjectionNoiseBackAction}

\newcommand{\gf}{g_F}
\newcommand{\gj}{g_J}

Any spin ensemble will have intrinsic uncertainty in its net spin ${\bf J}$, as a consequence of spin uncertainty relations. If the ensemble spins are associated to a magnetic moment, the magnetic field produced by the ensemble will similarly have an intrinsic uncertainty.  The possibility that this uncertain field acts back on the ensemble and introduces a self-interaction noise has at times been discussed in the OPM community \cite{LeeDAMOP2008, KitchingPC2012, LeePC2019}.  Here we consider a simple model of this hypothetical noise source. 

We consider a spherical region occupied by $N \gg 1$ spin-$1/2$ atoms with a total spin $J = N/2$, initially polarized along the $z$ axis, so that $\langle J_z \rangle \approx N/2$.  We note that the $J_x$ and $J_y$ components of the total spin vector obey the uncertainty relation $\delta J_x \delta J_y \ge |\langle J_z \rangle|/2 \approx N/4$.  Considering precession about the $y$ axis for a time $T$, as a measure of the field component $B_y$, we can associate the uncertainty  $\delta J_x$ with an angular uncertainty $\delta \theta = \delta J_y/\langle J_z \rangle$, and thus a magnetic field uncertainty 
\be
\delta B_{(\rm SPN)} = \frac{\delta J_x}{\langle J_z \rangle \gamma T} \ge \frac{2}{\gamma T \delta J_y },
\ee
where $\gamma = \gj \mu_B$ is the gyromagnetic ratio, and SPN indicates spin projection noise.  

 $\delta J_y$ can also be associated to a field uncertainty, through the magnetic field generated by the ensemble of spins. Modeling the ensemble as a uniformly-magnetized sphere, we find the self-generated field inside the sphere is ${\bf B} = 2 \mu_0 {\bf M}/3$, where ${\bf M} = \hbar \gamma {\bf J}/V$ is the magnetization density of the material \cite{GriffithsBook1999}. As we are interested in just the $B_y$ component, we have
\be
\label{eq:SelfDipoleField}
\delta B_{(\rm MSI)} =  \frac{2 \hbar \gamma \mu_0  \delta J_y}{3 V},
\ee
where MSI indicates magnetic self-interaction. 

If we suppose that these two noise sources are independent, the energy resolution per bandwidth is 
\bea
E_R & = & \frac{\langle \delta B^2 \rangle VT }{2 \mu_0}
= \frac{\langle \delta B_{(\rm SPN)}^2\rangle  +  \langle \delta B_{(\rm MSI)}^2\rangle}{2\mu_0}  VT
\nonumber \\ & \ge &
\frac{{C^2}{x^{-1}} + D^2 x}{2\mu_0} 
\ge \frac{CD}{\mu_0} = \frac{4}{3} \hbar,
\eea
where $C \equiv 2/\gamma $, $D\equiv 2 \hbar \gamma \mu_0/3$ and $x \equiv T \langle \delta J_y^2\rangle/V$.  The first inequality is saturated for minimum uncertainty states, i.e. those with $\delta J_x \delta J_y = |\langle J_z \rangle|/2$, and the second can be saturated by choosing $T$ such that $C^2 x^{-1} = D^2 x$.  

Once again, the result is consistent with observed values of $E_R$.  And once again, we believe the calculation is subtly misleading and does not in fact represent a limit.  For one thing, the precession angle, and thus $B_y$, can be inferred from a measurement of a single spin component, e.g. $J_x$ if the precession angle is small.  The noise in $J_x$ and the noise in $J_y$, if it contributes to the rotation speed, will contribute in linear combination to the field estimate. There will be spin-squeezed states that have  small uncertainty of this linear combination, producing a total noise $\langle\delta B^2\rangle$ far smaller than what is derived above. In few words, the sum in quadrature is not appropriate if $J_x$ and $J_y$ are correlated.  

A still more serious objection is that magnetic self-action of the kind assumed here, like the vacuum field effects described in Section \ref{sec:ZeroPoint}, appears unphysical.  The above model suggests that, in the absence of any external field, a spin system could reorient itself to have a net angular momentum different from its initial angular momentum.  Indeed, given enough time, it would sample all possible orientations.  This kind of `bootstrapping,' in which the spin system rotates itself does not conserve angular momentum (we note the spin system has no neighbors with which to exchange angular momentum).  The possibility that the angular momentum is taken up by the electromagnetic field suggests itself, but this would seem to violate energy conservation, as the spin system would forever radiate a fluctuating field.

\subsection{Margolus-Levitin bound} 
If we consider the magnetometer and field together as a single quantum system, the Margolus-Levitin (ML) theorem \cite{MargolusPFWPC1998} shows that this system cannot change from an initial state $|\psi_i\rangle$ to an orthogonal final state $|\psi_f\rangle$ in a time shorter than 
\begin{equation}
T_{\rm min} = \frac{ \pi \hbar}{2 E},
\end{equation}
where $E \equiv \langle H \rangle - E_0$ is the mean available energy,  $H$ is the Hamiltonian and $E_0$ is its lowest eigenvalue. If we identify $E$ with $B_{}^2 V/(2 \mu_0)$, i.e., the magnetic field energy within the sensor volume $V$, we can identify a minimum field strength $B_{\rm min}$ that produces orthogonality in time $T_{\rm min}$, and thus can reliably be distinguished from zero field.  This gives
\begin{equation}
\label{eq:MLbound}
\frac{B_{\rm min}^2 V  T_{\rm min}}{2 \mu_0} \ge  \frac{\pi \hbar}{2},
\end{equation}
which resembles an ERL, with the difference that the squared field appears, rather than the mean squared error of the field estimate.  

 As in with the energy-time uncertainty relation discussed above, this difference is important: If we consider  a small perturbation $\delta B$ to a large field $B_0$, it would be natural to divide the total energy into a fixed contribution $E_0 = B_{0}^2 V/(2 \mu_0)$ and the perturbation to the energy $\langle H \rangle \approx 2  \delta B  B_{0}V /(2 \mu_0)$, leading to 
\begin{equation}
\label{eq:MLboundP}
\frac{\delta B_{} B_0 V  T_{\rm min}}{\mu_0} \ge  \frac{\pi \hbar}{2},
\end{equation}
which for large $B_0$ allows detection of $\delta B \ll B_{\rm min}$.  Another serious concern for use of the ML bound to derive an ERL is the role of field-matter coupling.  For example, a sensor with magnetic moment ${\bf \mu}$ would contribute $-{\bf \mu} \cdot {\bf B}$ to the total energy.  If this is included in $E$ and ${\bf \mu}$ is allowed to become large, the minimum detectable field can vanish.  

\subsection{Bremermann-Bekenstein bound} In the context of black hole thermodynamics \cite{WaldPRD1979}, it was shown by Bremermann and Bekenstein (BB) \cite{BremermannIJTP1982,BekensteinPRD1981} that a spherical region ${\cal R}$ with radius $R$ contains a bounded information entropy 
\be
\label{eq:BBBound}
H_{\rm BB} \le \frac{2 \pi E R}{\hbar c},
\ee
where $E$ is the mean energy contained in the sphere. The result applies also to non-relativistic scenarios \cite{SchifferPRA1991} and has been applied to the question of the energy cost of communication \cite{BekensteinPRL1981, Bekenstein1984} by considering moving packets of material and/or radiation, echoing earlier bounds based on Shannon information capacity and energy-time uncertainty \cite{BremermannIJTP1982}.  

We may consider the read-out of the sensor as communication from the sensor to some other system, which might be a display, recording device, or an interested scientist.  As for the ML bound, it is natural to consider the magnetostatic field energy $E = \langle B^2\rangle V/(2\mu_0)$.  The entropy we interpret as an upper bound on the number of resolvable field states. One bit of message, corresponding to the minimum detectable field $B_{\rm min}$, is then achieved for a field energy
\begin{equation}
\frac{2 \pi \langle B_{\rm min}^2\rangle V R}{2\mu_0  c} \ge \hbar.
\end{equation}
We note that at $c$, the maximum speed of communication, information requires at least $\MeasDur = R/c$ to reach a single point from the entirety of ${\cal R}$.  Inserting into the above we find
\begin{equation}
\frac{\langle B_{\rm min}^2 \rangle V \MeasDur}{2\mu_0} \ge \frac{1}{2 \pi} \hbar,
\end{equation}
which resembles an ERL, with the difference that the mean squared field appears, rather than the mean squared error of the field estimate.  As with the ML bound, this appears to allow resolution of small increments on large fields. 

The nonlinearity implicit in Eq.~(\ref{eq:BBBound}), in which $H_{\rm BB}$ is both the logarithm of the number of possible states and proportional to the mean squared field, amplifies this concern.  We take as a reference a spherical region ${\cal R}$ containing a field $B$ with $\langle B^2 \rangle = B_{\rm min}^2$, sufficient to encode one bit of information, or equivalently to distinguish between two possible field states.  If we now imagine the same region containing a stronger field, with $\langle B^2 \rangle = \beta^2 B_{\rm min}^2$ for some $\beta > 1$,  Eq.~(\ref{eq:BBBound}) limits the entropy to $H_{\rm BB} = \beta^2 $ \SI{}{bits}, which can encode up to $2^{\beta^2}$ distinct states, distributed over the $\sim \beta B_{\rm min}$ range of the field distribution.  The minimum resolvable field increment $\delta B$ is then 
\be
\frac{\delta B}{B_{\rm min}} \sim \sqrt{\frac{\langle B^2 \rangle }{B_{\rm min}^2}}   \exp\left[- \frac{\langle B^2 \rangle }{B_{\rm min}^2} \right].
\ee
This describes an exponentially small minimum field increment, achieved when measuring large (or potentially large) fields.

\section{Systems proposed to surpass  $E_R = \hbar$}
\label{sec:ProposedToBeat}

In the preceding sections we have described both established technology-specific and potential technology-spanning quantum limits on the energy resolution.  No convincing technology-spanning limit was found, however, leaving open the possibility of sensing with unconstrained energy resolution.  In this section we describe sensing methods, both proposed and implemented, that appear to evade the technology-specific quantum limits presented above.

\subsection{Non-dissipative superconducting sensors} 

The TC limit arises due to the zero-point current fluctuations in the shunt resistances, the only dissipative components of the dcSQUID model analyzed by TC.  A sufficiently small shunt resistance prevents hysteresis, making the SQUID current a single-valued function of the flux to be detected.  The intrinsic noise of the dcSQUID could, within this model, vanish if the resistance were made infinite. The interpretation of the current signal would, however, be more complex.
Superconducting field sensors that do not include a dissipative element include superconducting quantum interference proximity transistors (SQUIPTs) \cite{GiazottoNP2010} and superconducting kinetic inductance magnetometers  \cite{KherEUCAS2013, LuomahaaraNC2014, KherJLTP2016, KherThesis2017}.  

\subsection{Localized single quantum systems} 
Single quantum systems (SQSs) such as NV centers \cite{TaylorNP2008} and single trapped ions \cite{BaumgartPRL2016} have been proposed as extremely high-spatial-resolution field sensors. Single Rydberg atoms have also been studied as magnetic sensors \cite{DietscheNPhys2019}.  Because they are elementary systems, internal decoherence mechanisms such as are described in Sections~\ref{sec:OPMLimit} and \ref{sec:NVDLimit} can be fully evaded.  These sensors are also potentially  small, with the effective linear dimension limited by the precision with which they can be localized and the minimum distance from possible sources.  It thus seems possible that SQSs would have no energy resolution limit. It is, nonetheless, worth noting that SQSs in solids experience a significant noise from surface effects \cite{MyersPRL2017}, an effect that becomes more important as effective linear dimension decreases.  Similarly, efforts to produce  small ion traps have uncovered important noise sources associated with closely-placed electrodes \cite{HiteMRSB2013}. 

\subsection{Dynamical decoupling and low-entropy reservoirs}
As described in Section \ref{sec:NVs}, fixed, spatially-disordered spin ensembles (e.g. NVD ensembles as currently implemented) experience a self-depolarization caused by the magnetic dipole-dipole coupling among elements of the ensemble. This depolarization can be understood as a transfer entropy from the center-of-mass (cm) degrees of freedom into the spin degrees of freedom by a coherent evolution that is not \textit{per se} a source of entropy.  

Two classes of methods present themselves to prevent such dipole-induced depolarization.  The first, more established method is dynamical decoupling \cite{ViolaPR1998}, in which strong, impulsive spin operations are applied to prevent or reverse the buildup of coherent rotations due to the naturally present dipolar coupling.  Such decoupling of sensing electrons from a surrounding bath of nuclear spins is an established method, e.g. with NVD \cite{PhamPRB2012}.  The application to an ensemble of sensor spins has the potential to dynamically decouple the spins from their neighbors, while leaving them coupled to the external field to be measured \cite{ChoiPRL2017}.  The theory and design of suitable pulse sequences  is an active topic \cite{AttarARX2019, HaasARX2019, ChoiARX2019}, and experimental results have been reported \cite{ZhouARX2019}.

The second approach is simply to remove the entropy from the reservoir, which could in principle be done by ordered positioning of the spins.  Similarly, phononic disorder can be reduced through cooling, although even at zero temperature the phononic vacuum presents a decoherence channel \cite{AstnerNM2018}.

Another possible route to entropy removal in spin-precession sensors is the use of quantum degenerate gases. In  such a system (a spinor BEC), two-body interactions, including both short-range ferro-/anti-ferromagnetic contact interactions and long-range dipole-dipole coupling, induce coherent spin evolutions rather than introducing entropy to the spins.  A jump in coherence lifetime at Bose-Einstein condensation has been observed in planar geometries \cite{HigbiePRL2005} and exploited for high-sensitivity BEC magnetic imagers  \cite{VengalattorePRL2007} and gradiometers \cite{VengalattorePRL2007, WoodPRA2015, JaspersePRA2017}.  A full ``freezing-out'' of the cm degrees of freedom has been observed in a quasi-zero-dimensional single-domain spinor BEC \cite{PalaciosNJP2018}.  These results were all obtained with the ferromagnetic ground state of $^{87}$Rb.

\subsection{Precessing ferromagnetic needle}
A similar ``freezing-out'' of non-spin degrees of freedom is predicted for  solid-state ferromagnets in the single-domain size regime \cite{JacksonKimballPRL2016}.  As with the BEC case described above, the ferromagnetic interactions impose full polarization, and at low temperatures no intrinsic fluctuations cause diffusion of the polarization angle.  Assuming background gas pressure, which imparts random angular momentum input, can be arbitrarily reduced, the sensitivity, limited by readout noise, is predicted to scale as $\langle \delta B^2 \rangle T \propto T^{-2}$. Thus for long measurements $E_R$ is predicted to have no lower limit in this system.  Variants based on free rotation \cite{JacksonKimballPRL2016} and on normal mode oscillations \cite{VinanteARX2019} have been proposed. 

\subsection{OPMs with low spin-destruction rates} 
\label{sec:SDRates}
As described in Section \ref{sec:OPMLimit}, the ERL arises in gas- and vapor-phase spin-precession magnetometers due to two-body relaxation processes, e.g. spin-destruction collisions, with a limit $E_R \ge  \bar{v} \sigma/(2 \mu_0 \gamma^2)$, where $\sigma$ is the relevant spin-relaxation cross section.  The figure of merit $\bar{v} \sigma \gamma^{-2}$ varies considerably among alkali and noble gas species, with lower values for noble gases and for lighter atoms \cite{NewburyPRA1993, KadlecekPRA2001, ChannPRL2002, BerryPuseyPRA2006}.  $\sigma\subSD$ for K is about an order of magnitude lower than for $^{87}$Rb \cite{AllredPRL2002}, such that the predicted ERL for a SERF-regime K vapor magnetometer is below $\hbar$.  Generalization about two-body relaxation is difficult, due to the coexistence of several mechanisms including transient dimer formation, spin-orbit coupling in second order, and magnetic dipole-dipole coupling \cite{KadlecekPRA2001}.  It is nonetheless instructive to consider the case of pure magnetic dipole-dipole relaxation, which appears to describe at least the case of gaseous $^3$He at large pressures \cite{GentileRMP2017}.  For this mechanism, the spin-destruction cross-section scales as $\sigma_{\rm dd} \propto \mu^4 \propto \gamma^4$ where $\mu$ is the atomic magnetic moment \cite{NewburyPRA1993}. As a result, the ERL figure of merit favors atomic species with small magnetic moments such as $^3$He.

\section{Summary and observations} 

In this \ArticleType we have reviewed the history and status of energy resolution limits in precise sensing of low-frequency magnetic fields.  We now recapitulate our findings and comment on their significance to ongoing efforts to improve sensor performance. First, we reviewed reported sensor performance, to find that the best reported sensors obey a limit $E_R \ge \hbar$, even as they span many orders of magnitude in size and field resolution.  We conclude that $E_R$ is, at a minimum, an interesting metric for comparing different technologies; it must in some meaningful way capture the challenges of achieving high sensitivity, speed and spatial resolution. The fact that the best achieved values for $E_R$ approach $\hbar$ is suggestive of a fundamental quantum limit.  

This suggestion is backed up by technology-specific ERLs, which are known for dc SQUIDs, alkali vapor OPMs, and fixed-position spin-precession sensors, e.g. NVD sensors.  These ERLs coincide in predicting a limit near $E_R = \hbar$.  The origins of these ERLs are quite technology-specific, involving shunt resistances, spin-destruction collisions, and random dipole-dipole couplings, respectively, and have not yet been brought together under any unifying principle. We have reviewed several  general quantum limits and their potential to supply a unifying, technology-spanning ERL. We in fact found several arguments that could be made to predict a technology-spanning ERL near $\hbar$. In our analysis, however, each of these arguments has some important weakness, and none of them convincingly implies an ERL.  Finally, we  reviewed proposed new sensor types, and modifications to existing sensor methodologies, which appear to escape all of the known ERLs and thus may provide $E_R$ values far below $\hbar$. Such sensors, if their current analyses are correct, have the potential to surpass today's leading magnetometer technologies.  

 The technology-specific ERLs each in some way concern dissipation mechanisms that are closely linked to the sensor's response to the applied field. In the case of dc SQUIDs, dissipation in the form of finite shunt resistance is introduced to achieve a single-valued steady-state response, i.e. to remove hysteresis. In alkali-vapor OPMs, the rate of alkali-induced spin-destruction collisions is proportional to the alkali number density, which directly impacts the projection-noise-limited signal-to-noise-ratio. In color-center sensors, the dissipation is similarly linked to magnetic dipole-dipole coupling, and has a similar dependence on number density.  Viewed as a group, they can be summarized by the proposition: a useful coupling of the sensor to the field of interest necessarily creates also a dissipation strong enough to impose an ERL.  In light of this, one can imagine a technology-spanning limit emerging from the theory of open quantum systems \cite{BreuerBook2007, DaviesBook1976, IngardenBook1997, LindbladBook2001, RotterRPP2015}. To our knowledge the question of ERLs has not been explored in that context. 

The proposals for new, ERL-surpassing sensors for the most part aim to alter an existing sensor methodology in such a way that it evades the above-mentioned mechanisms that link field response to dissipation. For example, single quantum systems are predicted to retain sensitivity to external fields while evading completely dipole-dipole coupling of sensor components, simply because there is only one component. More generally, the specificity of the known ERLs makes it plausible that a sensing system could be designed to evade them.  We thus find ourselves, at the end of this review, of two minds. On the one hand, the coincidence of multiple technology-specific quantum limits with each other, and with the empirical results of the most advanced sensor systems, makes it difficult to believe that there is not some as-yet-undiscovered general principle imposing ERLs on field sensors.  At the same time, we do not see any fundamental impediment to sensors with arbitrarily small $E_R$, if they are constructed to evade  the existing limits.  Perhaps the resolution to this dilemma is equally bifurcated: it may be that a broad class of sensors is subject to a yet-to-be-discovered ERL, while a second class, operating by other principles, escapes it.  We hope our observations in this \ArticleType will help to resolve this and other open questions in the topic of sensor energy resolution limits, and will ultimately help to advance sensor technology.

\section{Acknowledgements}

We thank J. Kitching for suggesting this problem, and for sharing his insights on several aspects of it, including unpublished work.  M. Romalis and S.-K. Lee shared unpublished work on limits set by spin-destruction collisions and spin-noise self-interaction.  We thank M. Lukin, I. Chuang,  and J. Hassel for helpful discussions regarding NVD and superconducting sensors. We thank R. J. Sewell for feedback on the manuscript.  Work supported by European Research Council (ERC) projects AQUMET (280169) and ERIDIAN (713682); European Union projects QUIC (Grant Agreement no.~641122) and FET Innovation Launchpad UVALITH (800901);~the Spanish MINECO projects OCARINA (Grant Ref. PGC2018-097056-B-I00) and Q-CLOCKS (PCI2018-092973), the Severo Ochoa programme (SEV-2015-0522); Ag\`{e}ncia de Gesti\'{o} d'Ajuts Universitaris i de Recerca (AGAUR) project (2017-SGR-1354); Fundaci\'{o} Privada Cellex and Generalitat de Catalunya (CERCA program, RIS3CAT project QuantumCAT); Quantum Technology Flagship projects MACQSIMAL (820393) and QRANGE (820405); Marie Sk{\l{odowska-Curie ITN ZULF-NMR (766402); EMPIR project USOQS (17FUN03). 

 \bibliography{./MegaBib}



\renewcommand{\bra}[1]{\langle #1|}
\renewcommand{\ket}[1]{|#1\rangle}

\appendix

\section{Thermal and zero-point magnetic noise}
\label{app:ZeroPoint}

We imagine a device that instantaneously makes an ideal measurement of the $z$ component of the magnetic field, within a spherical region ${\cal R}$ of volume $V_{\rm S} = 4 \pi \rSens^3/3$ where $\rSens$ is the radius of the region, which for convenience we take to be centred on the origin.

We describe this via the scalar observable
\begin{eqnarray}
\bar{B}_z(t) &\equiv&  \int d^3 r \rho(\br) \hat{z}\cdot {\bf B}(\br,t) 
\end{eqnarray}
where ${\bf B}(\br,t)$ is the quantized magnetic field. The weighting function $\rho(\br) \ge 0$ should be normalized $\int d^3 r \rho(\br) = 1$ and should vanish for $r > \rSens$, but is otherwise arbitrary. In what follows we use 
\be
\label{eq:WeightingFunction}
\rho(\br) = \frac{5}{2 V_{\rm S}} \left\{
\begin{array}{lr}
1 - r^2/\rSens^2 & r \le \rSens \\
0 & r > \rSens
\end{array} \right. , 
\ee
which gives relatively simple results. 
 The quantized magnetic field is
\begin{eqnarray}
{\bf B}(\br,t)&\equiv& {\bf B}^{(+)}(\br,t)  + {\bf B}^{(-)}(\br,t) 
\end{eqnarray}
where
\begin{eqnarray}
{\bf B}^{(+)}(\br,t) &\equiv& i \sum_{\bk,\alpha} \sqrt{\frac{\mu_0 \hbar \omega_k}{2 L^3}}   \bff_{\bk, \alpha} a_{\bk,\alpha} e^{i \bk \cdot \br - i \omega_{\bk} t} \\
{\bf B}^{(-)}(\br,t) &\equiv& \left[ {\bf B}^{(+)}(\br,t)  \right]^\dagger,
\end{eqnarray} 
$\omega_k = c |k|$, ${\bf f}_{{\bf k},\alpha}$ is a unit vector describing the polarization of mode $({\bf k}, \alpha)$ with annihilation operator $a_{{\bf k},\alpha}$, and $L$ is the side-length of the quantization volume, later taken to infinity.
We similarly define $\bar{B}_z^{(+)}(t) \equiv  \int d^3r\, \rho(\br) \, \hat{z}\cdot {\bf B}^{(+)}(\br,t)$ and 
$\bar{B}_z^{(-)}(t) \equiv [\bar{B}_z^{(+)}(t)]^\dagger$.

\newcommand{\krsSymb}{\zeta}

We can then use $\langle (a a^\dagger + a^\dagger a)/2 \rangle = \langle a^\dagger a \rangle + 1/2 = \langle n \rangle + 1/2$, which for the thermal state of the field at temperature $T_B$ has a value $1/2 + (\exp[\hbar \omega/k_B T_B]-1) \approx 1/2 + k_B T_B / \hbar \omega$ when $k_B T_B \gg \hbar \omega$ (Rayleigh-Jeans law), to obtain
\begin{eqnarray}
\langle \bar{B}_z^2 \rangle & = & \langle \bar{B}_z^{(-)}\bar{B}_z^{(+)}  +  \bar{B}_z^{(+)}\bar{B}_z^{(-)} \rangle  
\nne \frac{\mu_0 \hbar c}{L^3} \sum_{\bk,\alpha} (\frac{1}{2} + \frac{k_B T_B}{\hbar c k}) k |\bff_{\bk, \alpha} \cdot \hat{z}|^2 
\left| \int d^3r\, \rho(\br) e^{i \bk \cdot \br} \right|^2 \nn
\end{eqnarray}
The integral over $\br$ we compute in spherical polar coordinates with the polar axis along $\bk$:
\bea
 \int d^3r\, \rho(\br) e^{i \bk \cdot \br } &=& \frac{2 \pi}{V} \int_0^\pi \sin\theta_\rho d\theta_\rho \int_0^\rSens r^2 dr  \, e^{i k r \cos \theta_\rho } 
 \nne
 \frac{20 \pi  \left[ \left(3- \krsSymb^2 \right) \sin (\krsSymb)- 3 \krsSymb \cos (\krsSymb)\right]}{k^3 \krsSymb^2 V}, \hspace{8mm}
\eea
where $\krsSymb \equiv k \rSens$. 

Now using spherical polar coordinates in which the polar axis is along $\hat{z}$, so that  $\bk = k(\sin\theta\cos\phi,\sin\theta\sin\phi,\cos\theta)$, and choosing polarization modes $\bff_{\bk,\alpha}$ so that one is orthogonal to  $\hat{z}$, the other has $\bff_{\bk,\alpha} \cdot \hat{z} = \sin\theta$. 
Using the density of states $L^3/(2\pi)^3$, we have 
\begin{eqnarray}
 \sum_{\bk,\alpha} |\bff_{\bk, \alpha} \cdot \hat{z}|^2 &\rightarrow& \frac{L^3}{2^3\pi^3} \int k^2 dk\, \sin\theta\, d\theta\, d\phi\,  \sin^2\theta
 \nne \frac{L^3}{3\pi^2}  \int k^{2}\,  dk.
\end{eqnarray}

Combining the above we find
\begin{eqnarray}
\langle \bar{B}_z^2 \rangle & = &  \frac{\mu_0 \hbar c}{L^3}\frac{L^3}{3\pi^2}  
 \int_0^\infty dk \, (\frac{1}{2} k^3 + \frac{k_B T_B}{\hbar c} k^2) 
 \nnt 
\left|  
\frac{20 \pi  \left[ \left(3-\krsSymb^2\right) \sin (\krsSymb)-3 \krsSymb \cos (\krsSymb)\right]}{k^3 \krsSymb^2 V}
\right|^2
\nne
\frac{25  c \mu_0}{8 \pi^2 \rSens^4}\hbar + \frac{5 \mu_0}{7 \pi \rSens^3} k_B T_B.
\eea

This gives the sensitivity of a single instantaneous measurement.  To avoid measurement back-action it is necessary to wait a finite time before making the next measurement:  the measurement will introduce noise into the observable conjugate to $\bar{B}_z(t)$, and by Maxwell's equations this disturbance will propagate into $\bar{B}_z(t'>t)$.  The disturbance will propagates fully outside of ${\cal R}$  (it will always propagate at $c$) in a time $\MeasDur = 2 \rSens/c$, enabling back-action-free repeated measurements with repeat period $\MeasDur$.

The resulting energy resolution per bandwidth is
\begin{eqnarray}
 E_R & = & \frac{\langle\bar{B}_z^2 \rangle V \MeasDur }{ 2\mu_0} 
\nne
\frac{175}{42 \pi} \hbar + \frac{20 \rSens }{21 c}   k_B T_B.
\end{eqnarray}
The first term describes the quantum noise contribution to the measurement.  The pre-factor (here $175/(42 \pi) \approx 1.3$) depends on the precise choice of weighting function $\rho(\br)$.

\newcommand{\ROp}[1]{R_{#1}({\bf n},\theta)}
\newcommand{\ROpS}{\ROp{\rm S}}
\newcommand{\ROpF}{\ROp{\rm F}}

\section{Zero-point magnetic noise and spin-precession sensors}
\label{sec:QuantumSpinVacuum}

A paradigmatic field-sensing protocol with spin systems consists of preparing a magnetic spin system in a known direction, allowing it to interact with the field for a known free-precession time $T$, and detecting the spin orientation by a projective measurement.  Here we consider the role of zero-point fluctuations of the field in this protocol.  

We first consider this scenario semi-classically. It is sufficient to consider a Hilbert space ${\cal H} = {\cal H}_{\rm M}$ describing the material system (here and below we use the subscripts $_{\rm M}$ and $_{\rm F}$ to indicate material and field, respectively) with dynamics governed by the Hamiltonian 
\be
\label{eq:SemiclassHamiltonian}
H_{\rm SC} = H_{\rm M} + H_{\rm MF}.
\ee
Here the spin-field interaction is $H_{\rm MF} = -\mu \cdot {\bf B} = - \gamma \hbar {\bf S} \cdot {\bf B}$,  where ${\bf S}$ is the net spin and $\gamma$ is the gyromagnetic ratio.   In this semiclassical description ${\bf B}$ is a c-number field, and is unchanged by the interaction with the material system. 

For the protocol to be efficient, it should be possible to prepare a stable, strongly polarized state of the material system, and this state should be free to rotate. We thus assume the following properties of $H_{\rm M}$: 1) The ground state is continuously degenerate under rotations, such that any state of the form $\ROpS \ket{\psi_0}_{\rm M}$ is a ground state, where $\ROpS$ rotates all spins about an axis ${\bf n}$ by angle $\theta$. 2) The reference ground state $\ket{\psi_0}_{\rm M}$ is polarized: $S_z \ket{\psi_0}_{\rm M} = S \ket{\psi_0}_{\rm M}$. 3) The ground states are separated from any non-spin-$S$ states by an energy gap.  4) For simplicity, we assume decoupling of spin and centre-of-mass degrees of freedom. Unlike a compass needle, here a spin rotation does not imply a rotation of the mass of the system. These assumptions describe an isotropic ferromagnet, but also a single electron, an atom with a non-spin-zero ground state, or a single-domain spinor condensate.  

Starting from any ground state of this system, evolution under a small classical B-field simply rotates within the space of ground states of $H_{\rm M}$, with no addition of energy or entropy. The intrinsic angular noise in the sensing protocol is then independent of $T$, leading to a variance of the inferred field $\langle \delta B^2 \rangle\propto T^{-2}$ and $E_R \propto T^{-1} \rightarrow 0$ for large $T$.

To consider this protocol with a quantized field, we first expand the Hilbert space and Hamiltonian to include the field: ${\cal H}_{\rm MF} \equiv {\cal H}_{\rm M} \otimes {\cal H}_{\rm F}$ and 
\be
\label{eq:MFHamiltonian}
H = H_{\rm M}\otimes \mathbb{I}_{\rm F} + \mathbb{I}_{\rm M} \otimes H_{\rm F} + H_{\rm MF}.
\ee
Here $H_{\rm F} = \int d^3 x \, (  \epsilon_0{\bf E} \cdot {\bf E}/2 + {\bf B} \cdot {\bf B}/2\mu_0) $ is the field Hamiltonian.  ${\bf E}$ and ${\bf B}$ are quantized fields. The ground state of $H_{\rm F}$ is the vacuum state $\ket{0}_{\rm F}$. 

\newcommand{\Bbar}{\bar{\bf B}}
\newcommand{\subBbar}{_{\Bbar}}

For simplicity, we consider using the sensor to estimate an externally applied field that is constant in time and uniformly distributed, making a contribution $\Bbar$ (a c-number vector) to the value of the field ${\bf B}({\bf x})$.  Such a contribution is described by a displacement operator $D\subBbar$, defined such that $D^\dagger\subBbar {\bf B}({\bf x}, t)  D\subBbar  = {\bf B}({\bf x}, t) + \Bbar$. We note that  the remaining field $\tilde{\bf B}({\bf x}, t) \equiv {\bf B}({\bf x}) - \Bbar$ describes both the zero-point field fluctuations and any field produced by the sensor itself.  We can consider the effect of this external field on the system as a whole by displacing the $H$ of Eq.~(\ref{eq:MFHamiltonian}) to find 
\bea
\label{eq:MFHamiltonianDisp}
D^\dagger\subBbar H D\subBbar  &=& H_{\rm M}\otimes \mathbb{I}_{\rm F} + \mathbb{I}_{\rm M} \otimes H_{\rm F} + H_{\rm MF} 
\nnp \frac{L^3}{2\mu_0} |\Bbar|^2   \mathbb{I}_{\rm M} \otimes  \mathbb{I}_{\rm F} - \gamma \hbar \Bbar \cdot {\bf S}  \otimes  \mathbb{I}_{\rm F},
\eea
where $ \mathbb{I}_{\rm M}$ and $\mathbb{I}_{\rm F}$ indicate identity operators and $L^3$ is the volume of the calculation. The last two terms describe the classical energy of the external field $\Bbar$ and the interaction of the spin with the external field, respectively.  This last contribution produces a torque $\gamma {\bf S} \times \Bbar$  on ${\bf S}$, and thereby induces a precession of ${\bf S}$ about $\Bbar$. 

The dynamics of ${\bf S}$ in this full-quantized model will depend on what one takes to be the initial state.  If one takes $\ket{\psi_0}_{\rm M}\otimes \ket{0}_{\rm F}$, i.e., a product state of the material and vacuum, the state will evolve, as it is not an eigenstate of $H$, which contains $H_{\rm MF}$.  Given the ``white noise'' character of vacuum fluctuations, it is not implausible that this evolution would produce a diffusion of ${\bf S}$, with an angular variance $\propto T$,  a variance of the field estimate $\langle \delta B^2 \rangle\propto T^{-1}$ and thus $E_R \propto T^{0}$, which describes an ERL.  An estimate of the limiting value of $E_R$  is given in \autoref{sec:ZeroPoint}.  

It is not clear that $\ket{\psi_0}_{\rm M}\otimes \ket{0}_{\rm F}$ is a realistic starting condition, however.  It describes a scenario in which the spin ${\bf S}$ is oriented along a specific direction but the field around it is on average zero. It thus describes a condition in which the spin has not (or not yet) produced a dipolar field around itself.  This ``bare spin'' condition is one that we don't expect to arise naturally in any experiment.  And even if a bare spin could be made in some way, it surely would not remain bare for long.  

A more reasonable starting condition is the ``dressed'' version of $\ket{\psi_0}_{\rm M}$, which is to say a ground state of $H$ with the same symmetry under rotations as $\ket{\psi_0}_{\rm M}\otimes \ket{0}_{\rm F}$. In such a state, the field ${\bf B}({\bf x})$ will presumably contain, in addition to vacuum fluctuations, the dipole field pattern produced by the magnetic moment $\mu$.  We note that $H$ and its components are all invariant under rotations of the coordinate system, or equivalently of ${\bf S}$, ${\bf B}({\bf x})$ and ${\bf x}$ together. The ground state of $H_{\rm F}$ is irrotational or spin-0 and the ground state of $H_{\rm M}$ is (by assumption) of spin $S$.  When these subsystems are perturbatively coupled by the irrotational $H_{\rm MF}$, $\ket{\psi_0}_{\rm M}\otimes \ket{0}_{\rm F}$ becomes a dressed ground state  $\ket{\Psi_0}_{\rm MF}$ with the same symmetry under global rotations\footnote{We assume here that a perturbative treatment is appropriate. A sufficiently strong coupling could in principle take the system through a quantum phase transition, with the result that the ground states of  $H$ have different symmetry than the ground states of $H_{\rm M} + H_{\rm F}$. This does not seem to be the case in quantum electrodynamics, in which dressed spins have the same symmetries as bare spins, and the coupling manifests in perturbative modifications to the spin properties, e.g. to the magnetic moments.}. In the same way, the other bare ground states $\ROpS\ket{\psi_0}_{\rm M}\otimes \ket{0}_{\rm F}$ give rise to spin- and field-rotated states $\ROpS \otimes \ROpF \ket{\Psi_0}_{\rm MF}$, where $\ROpF$ is the field rotation operator.   As ground states of the total Hamiltonian, they do not evolve in any way for $\Bbar = {\bf 0}$, and the intrinsic spin uncertainty is independent of $T$.

We note that $\ket{\Psi_0}_{\rm MF}$ is an entangled state of the material and field, and that the field must carry some of the angular momentum.  A projective measurement on the ``bare'' spin (if such a measurement could be made) would then have an additional intrinsic uncertainty, owing to the fact that the material spin system is one component of an entangled state.  In contrast to the hypothetical diffusion of the spin orientation described above, this would be a one-time noise contribution, in effect a contribution to the intrinsic spin noise of the state. Such a contribution would not change the scaling of $E_R$ with $T$.  

For an applied field $\Bbar \ne {\bf 0}$, the torque on ${\bf S}$ will cause precession of the spin and a matched rotation of the field it generates via $H_{\rm MF}$.  In contrast to the semiclassical case, the system will not remain exactly within the the manifold of dressed states. Rather, the precessing spin presents an oscillating magnetic dipole and will radiate, whereas the dressed ground states $\ROpS \otimes \ROpF \ket{\Psi_0}_{\rm MF}$ have no radiating component. Via radiation the spin will lose energy and eventually align with or against $\Bbar$, depending on the sign of $\gamma$. The rate of radiative relaxation scales as $\Bbar^3$, so that small fields could still be measured with very small $E_R$ by the simple protocol considered thus far.   We conclude that zero-point fluctuations do not in fact set a limit to $E_R$ in a spin precession measurement. 

\newcommand{\sBbar}{\bar{B}}

Finally, we note that an only slightly more complex protocol achieves $E_R \rightarrow 0$ also for strong fields. There will be some maximum degree of relaxation for which the state remains effectively within the dressed ground state manifold and the relaxation-free scaling $\langle \delta B^2 \rangle \propto T^{-2}$ holds. This implies a maximum free-precession time $\propto \sBbar^{-3}$, where $\sBbar = |\Bbar|$. A measurement made with this value of $T$ has sensitivity $\langle \delta B^2 \rangle \propto \sBbar^{6}$.  Consider now a closed-loop measurement protocol in which $\sBbar = {B}_{\rm unknown} +  {B}_{\rm control}$,  where ${B}_{\rm control}$ is a known control field and  ${B}_{\rm unknown}$ is an unknown field to be estimated. The protocol uses a sequence of measurements of $\sBbar$ and feedback to ${B}_{\rm control}$ to set $\sBbar$ closer and closer to zero. The unknown field is inferred from the value of the control field that produces this condition.  

  Assuming the feedback procedure is limited by the intrinsic measurement variance $\langle \delta B^2 \rangle \propto \sBbar^{6}$, then after the $n$th measurement, with variance  $\langle \delta B^2 \rangle_{n}$, feedback achieves $\langle \sBbar \rangle = 0$ and $\langle \delta \sBbar^2 \rangle_n =  \langle \delta B^2 \rangle_{n}$. If the free-precession time for the next measurement is then chosen $T_{n+1} \propto  \langle \delta \sBbar^2 \rangle^{-3/2}_{n}$ to ensure small relaxation, we find $T_{n+1} \propto  T_{n}^3$ and thus $\langle \delta \sBbar^2 \rangle_{n+1} \propto \langle \delta \sBbar^2 \rangle_n^3$. Assuming the first measurement could be made with sufficient precision to reduce the variance through the feeedback process, $\langle \delta \sBbar^2 \rangle_n$ thus decreases very rapidly with $n$.  The total time used in the protocol is dominated by the last measurement, and $E_R \rightarrow 0$.

\clearpage
\section{Literature values}
\label{sec:LiteratureData}
\begin{widetext}
\begin{table}[h!]
\begin{center}
\begin{tabular}{cp{2cm}ccccccclc}
&&&$l_1$&$l_2$&$l_3$ &$V$ &$A$ &$\delta \Phi \sqrt{T}$ &$\delta B \sqrt{T}$  & Notes \\
Label &References \ &Type \ &(\SI{}{\meter})& (\SI{}{\meter})& (\SI{}{\meter})& (\SI{}{\meter\cubed})& (\SI{}{\meter\squared})&(\SI{}{\tesla\per\sqrt\hertz})&(\SI{}{\weber\per\sqrt\hertz}) &  \\
1&[1]&SQUID&\SI{3.0e-2}{}&\SI{3.4e-2}{}&$\cdot$&$\cdot$&\SI{9.9e-4}{}&$\cdot$&\SI{1.5e-16}{}& \\
2&[2]&OPM&\SI{5.0e-3}{}&\SI{5.0e-3}{}&\SI{1.8e-2}{}&\SI{4.5e-7}{}&$\cdot$&$\cdot$&\SI{1.6e-16}{}& \\
3&[3]&SQUID&\SI{4.5e-2}{}&\SI{4.5e-2}{}&$\cdot$&$\cdot$&\SI{1.6e-3}{}&$\cdot$&\SI{2.0e-16}{}& \\
4&[4]&SQUID&\SI{1.2e-2}{}&\SI{1.2e-2}{}&$\cdot$&$\cdot$&\SI{1.5e-4}{}&$\cdot$&\SI{3.3e-16}{}& \\
5&[5]&OPM&\SI{4.0e-2}{}&\SI{4.0e-3}{}&\SI{3.1e-3}{}&\SI{3.0e-7}{}&$\cdot$&$\cdot$&\SI{5.4e-16}{}& \\
6&[6]&OPM&$\cdot$&$\cdot$&$\cdot$&\SI{6.6e-7}{}&$\cdot$&$\cdot$&\SI{5.4e-16}{}& \\
7&[4]&SQUID&\SI{7.0e-3}{}&\SI{7.0e-3}{}&$\cdot$&$\cdot$&\SI{4.9e-5}{}&$\cdot$&\SI{7.0e-16}{}& \\
8&[7]&SQUID&\SI{7.3e-3}{}&\SI{7.3e-3}{}&$\cdot$&$\cdot$&\SI{4.4e-5}{}&$\cdot$&\SI{9.0e-16}{}& \\
9&[8,9,10]&SQUID&\SI{1.6e-2}{}&\SI{1.6e-2}{}&$\cdot$&$\cdot$&\SI{2.6e-4}{}&$\cdot$&\SI{3.5e-15}{}& \\
10&[11]&OPM&\SI{1.0e-3}{}&\SI{1.0e-3}{}&\SI{1.0e-3}{}&\SI{1.0e-9}{}&$\cdot$&$\cdot$&\SI{5.0e-15}{}& \\
11&[12,13]&GMR&\SI{4.4e-4}{}&\SI{4.4e-4}{}&$\cdot$&$\cdot$&\SI{1.9e-7}{}&$\cdot$&\SI{3.2e-14}{}& \\
12&[14]&SKIM&\SI{2.0e-2}{}&\SI{2.0e-2}{}&$\cdot$&$\cdot$&\SI{4.0e-4}{}&$\cdot$&\SI{3.2e-14}{}& \\
13&[15]&SQUID&\SI{3.0e-3}{}&\SI{3.0e-3}{}&$\cdot$&$\cdot$&\SI{9.0e-6}{}&$\cdot$&\SI{8.5e-14}{}& \\
14&[16]&SQUID&\SI{1.0e-3}{}&\SI{1.0e-3}{}&$\cdot$&$\cdot$&\SI{7.9e-7}{}&$\cdot$&\SI{1.5e-13}{}& \\
15&[16]&SQUID&\SI{5.0e-4}{}&\SI{5.0e-4}{}&$\cdot$&$\cdot$&\SI{2.0e-7}{}&$\cdot$&\SI{2.4e-13}{}& \\
16&[17]&SQUID&\SI{2.5e-5}{}&\SI{2.5e-5}{}&$\cdot$&$\cdot$&\SI{6.3e-10}{}&\SI{1.7e-22}{}&\SI{2.8e-13}{}& \\
17&[16]&SQUID&\SI{5.0e-4}{}&\SI{5.0e-4}{}&$\cdot$&$\cdot$&\SI{2.0e-7}{}&$\cdot$&\SI{3.3e-13}{}& \\
18&[18,19]&YIG&\SI{1.0e-2}{}&\SI{1.0e-2}{}&$\cdot$&$\cdot$&\SI{1.0e-4}{}&$\cdot$&\SI{4.0e-13}{}& \\
19&[16]&SQUID&\SI{2.5e-4}{}&\SI{2.5e-4}{}&$\cdot$&$\cdot$&\SI{4.9e-8}{}&$\cdot$&\SI{4.5e-13}{}& \\
20&[20]&BEC&\SI{1.1e-5}{}&\SI{1.1e-5}{}&$\cdot$&$\cdot$&\SI{1.2e-10}{}&$\cdot$&\SI{5.0e-13}{}& \\
21&[21]&CEOPM&\SI{2.8e-2}{}&\SI{2.8e-2}{}&\SI{2.8e-2}{}&\SI{1.1e-5}{}&$\cdot$&$\cdot$&\SI{6.9e-13}{}& \\
22&[16]&SQUID&\SI{2.5e-4}{}&\SI{2.5e-4}{}&$\cdot$&$\cdot$&\SI{4.9e-8}{}&$\cdot$&\SI{8.5e-13}{}& \\
23&[22]&RFNVD&\SI{9.5e-5}{}&\SI{9.5e-5}{}&\SI{9.5e-5}{}&\SI{8.5e-13}{}&$\cdot$&$\cdot$&\SI{9.0e-13}{}&See also [64]\\
24&[23]&FCNVD&\SI{1.0e-2}{}&\SI{1.0e-2}{}&\SI{2.0e-2}{}&\SI{2.6e-7}{}&$\cdot$&$\cdot$&\SI{9.0e-13}{}& \\
25&[16]&SQUID&\SI{4.0e-5}{}&\SI{4.0e-5}{}&$\cdot$&$\cdot$&\SI{1.6e-9}{}&$\cdot$&\SI{1.5e-12}{}& \\
26&[24]&SQUID&\SI{2.0e-4}{}&\SI{2.0e-4}{}&$\cdot$&$\cdot$&\SI{4.0e-8}{}&$\cdot$&\SI{1.7e-12}{}& \\
27&[25]&SQUID&\SI{4.0e-5}{}&\SI{3.5e-7}{}&$\cdot$&$\cdot$&\SI{1.4e-11}{}&\SI{3.6e-23}{}&\SI{2.6e-12}{}& \\
28&[26]&SQUID&\SI{2.0e-4}{}&\SI{2.0e-4}{}&$\cdot$&$\cdot$&\SI{4.0e-8}{}&$\cdot$&\SI{3.0e-12}{}& \\
29&[27,25]&SQUID&\SI{3.0e-5}{}&\SI{3.5e-7}{}&$\cdot$&$\cdot$&\SI{1.1e-11}{}&\SI{3.9e-23}{}&\SI{3.8e-12}{}& \\
30&[28]&BEC&\SI{4.7e-5}{}&\SI{2.1e-5}{}&\SI{2.1e-5}{}&\SI{2.0e-14}{}&$\cdot$&$\cdot$&\SI{3.9e-12}{}& \\
31&[29]&OPM&\SI{1.0e-3}{}&\SI{2.0e-3}{}&\SI{1.0e-3}{}&\SI{2.0e-9}{}&$\cdot$&$\cdot$&\SI{5.0e-12}{}& \\
32&[30]&COPM&\SI{2.0e-5}{}&\SI{2.0e-5}{}&\SI{3.0e-3}{}&\SI{3.7e-12}{}&$\cdot$&$\cdot$&\SI{5.4e-12}{}& \\
33&[31]&MFME&\SI{3.0e-2}{}&\SI{2.0e-3}{}&\SI{2.0e-4}{}&\SI{1.2e-8}{}&$\cdot$&$\cdot$&\SI{6.2e-12}{}& \\
34&[32]&SQUID&\SI{5.0e-6}{}&\SI{5.0e-6}{}&$\cdot$&$\cdot$&\SI{2.5e-11}{}&\SI{3.1e-22}{}&\SI{1.3e-11}{}& \\
35&[32]&SQUID&\SI{3.0e-6}{}&\SI{3.0e-6}{}&$\cdot$&$\cdot$&\SI{9.0e-12}{}&\SI{1.4e-22}{}&\SI{1.4e-11}{}& \\
36&[33]&BEC&$\cdot$&$\cdot$&$\cdot$&\SI{2.9e-14}{}&$\cdot$&$\cdot$&\SI{1.0e-11}{}& \\
37&[34]&BEC&\SI{1.0e-5}{}&\SI{10.0e-6}{}&$\cdot$&$\cdot$&\SI{1.0e-10}{}&$\cdot$&\SI{1.2e-11}{}& \\
38&[35]&NVD&\SI{1.3e-5}{}&\SI{2.0e-4}{}&\SI{2.0e-3}{}&\SI{5.2e-12}{}&$\cdot$&$\cdot$&\SI{1.5e-11}{}& \\
39&[32]&SQUID&\SI{1.0e-6}{}&\SI{1.0e-6}{}&$\cdot$&$\cdot$&\SI{1.0e-12}{}&\SI{9.3e-23}{}&\SI{3.6e-11}{}& \\
40&[36]&FCOPM&\SI{2.5e-4}{}&\SI{5.0e-4}{}&$\cdot$&$\cdot$&\SI{1.3e-7}{}&$\cdot$&\SI{2.3e-11}{}& \\
41&[37]&MFME&\SI{2.0e-4}{}&\SI{9.0e-4}{}&$\cdot$&$\cdot$&\SI{1.8e-7}{}&$\cdot$&\SI{2.7e-11}{}& \\
42&[38]&OPM&\SI{1.0e-3}{}&\SI{1.0e-3}{}&\SI{1.0e-3}{}&\SI{1.0e-9}{}&$\cdot$&$\cdot$&\SI{5.0e-11}{}& \\
43&[39]&RSC&\SI{5.0e-3}{}&\SI{2.8e-7}{}&$\cdot$&$\cdot$&\SI{1.4e-9}{}&$\cdot$&\SI{5.8e-11}{}& \\
44&[40]&BEC&\SI{1.1e-6}{}&\SI{1.1e-6}{}&\SI{4.0e-6}{}&\SI{2.0e-17}{}&$\cdot$&$\cdot$&\SI{7.7e-11}{}& \\
45&[18]&EMR&\SI{5.0e-5}{}&\SI{5.0e-5}{}&$\cdot$&$\cdot$&\SI{2.5e-9}{}&$\cdot$&\SI{1.0e-10}{}& 
\end{tabular}

\end{center}
\caption{Dimensions and field/flux sensitivities for the sensing results shown in Fig.~1.  References refer to the bibliography in Appendix~\ref{sec:References2}. Values are taken directly from the cited publications when possible. In some cases values are estimated, e.g. from a graph or an image.  Dots indicate values not found in the cited works.  \label{tab:SensSizeTable}
}
\end{table}
\newpage
\begin{table}[h!]
\begin{center}
\begin{tabular}{cp{2cm}ccccccclc}
&&&$l_1$&$l_2$&$l_3$ &$V$ &$A$ &$\delta \Phi \sqrt{T}$ &$\delta B \sqrt{T}$  & Notes \\
Label &References \ &Type \ &(\SI{}{\meter})& (\SI{}{\meter})& (\SI{}{\meter})& (\SI{}{\meter\cubed})& (\SI{}{\meter\squared})&(\SI{}{\tesla\per\sqrt\hertz})&(\SI{}{\weber\per\sqrt\hertz}) &  \\
46&[41]&SQUID&\SI{3.0e-6}{}&\SI{3.0e-6}{}&$\cdot$&$\cdot$&\SI{9.0e-12}{}&\SI{1.0e-21}{}&\SI{1.1e-10}{}& \\
47&[42]&PAFG&\SI{8.0e-3}{}&\SI{1.0e-3}{}&\SI{1.5e-5}{}&\SI{1.2e-10}{}&$\cdot$&$\cdot$&\SI{1.8e-10}{}& \\
48&[43,44,45]&BEC&\SI{3.0e-6}{}&\SI{3.0e-6}{}&\SI{3.0e-6}{}&\SI{2.7e-17}{}&$\cdot$&$\cdot$&\SI{2.3e-10}{}& \\
49&[46]&NVD&\SI{3.0e-3}{}&\SI{3.0e-3}{}&\SI{3.0e-4}{}&\SI{2.7e-9}{}&$\cdot$&$\cdot$&\SI{2.9e-10}{}& \\
50&[47]&GMR&\SI{9.0e-6}{}&\SI{3.6e-5}{}&$\cdot$&$\cdot$&\SI{3.2e-10}{}&$\cdot$&\SI{3.0e-10}{}& \\
51&[48]&COPM&\SI{5.0e-3}{}&\SI{2.0e-5}{}&\SI{2.0e-5}{}&\SI{2.0e-12}{}&$\cdot$&$\cdot$&\SI{3.2e-10}{}& \\
52&[49]&SQUIPT&\SI{1.1e-5}{}&\SI{1.1e-5}{}&$\cdot$&$\cdot$&\SI{1.2e-10}{}&\SI{4.1e-20}{}&\SI{3.4e-10}{}& \\
53&[50]&SQUID&\SI{3.3e-6}{}&\SI{6.7e-6}{}&$\cdot$&$\cdot$&\SI{2.2e-11}{}&\SI{1.1e-20}{}&\SI{5.1e-10}{}& \\
54&[51]&SQUID&\SI{2.0e-6}{}&\SI{2.0e-6}{}&$\cdot$&$\cdot$&\SI{3.1e-12}{}&\SI{4.1e-21}{}&\SI{1.3e-9}{}& \\
55&[52]&BEC&\SI{2.5e-4}{}&\SI{1.5e-6}{}&\SI{1.5e-6}{}&\SI{9.0e-17}{}&$\cdot$&$\cdot$&\SI{1.9e-9}{}& \\
56&[53]&NVD&\SI{1.8e-5}{}&\SI{1.8e-5}{}&\SI{3.1e-3}{}&\SI{3.5e-11}{}&$\cdot$&$\cdot$&\SI{3.0e-9}{}& \\
57&[54]&SQUID&\SI{6.5e-7}{}&\SI{6.5e-7}{}&$\cdot$&$\cdot$&\SI{4.2e-13}{}&\SI{1.8e-21}{}&\SI{4.3e-9}{}& \\
58&[55]&SQUID&\SI{1.6e-7}{}&\SI{1.6e-7}{}&$\cdot$&$\cdot$&\SI{2.0e-14}{}&\SI{1.0e-22}{}&\SI{5.1e-9}{}& \\
59&[56]&BEC&\SI{2.0e-6}{}&\SI{2.0e-6}{}&$\cdot$&$\cdot$&\SI{4.0e-12}{}&$\cdot$&\SI{6.0e-9}{}& \\
60&[57]&RFNVD&\SI{5.0e-7}{}&$\cdot$&$\cdot$&$\cdot$&$\cdot$&$\cdot$&\SI{3.8e-8}{}& \\
61&[55]&SQUID&\SI{5.6e-8}{}&\SI{5.6e-8}{}&$\cdot$&$\cdot$&\SI{2.5e-15}{}&\SI{1.0e-22}{}&\SI{4.2e-8}{}& \\
62&[58]&RFNVD&\SI{4.0e-9}{}&$\cdot$&$\cdot$&$\cdot$&$\cdot$&$\cdot$&\SI{5.3e-8}{}& \\
63&[59]&RFNVD&\SI{2.5e-8}{}&$\cdot$&$\cdot$&$\cdot$&$\cdot$&$\cdot$&\SI{5.6e-8}{}& \\
64&[55]&SQUID&\SI{4.6e-8}{}&\SI{4.6e-8}{}&$\cdot$&$\cdot$&\SI{1.7e-15}{}&\SI{1.0e-22}{}&\SI{6.2e-8}{}& \\
65&[60]&GRA&\SI{1.6e-4}{}&\SI{1.6e-4}{}&$\cdot$&$\cdot$&\SI{2.6e-8}{}&$\cdot$&\SI{1.0e-7}{}& \\
66&[61]&HALL&\SI{2.0e-7}{}&\SI{2.0e-7}{}&$\cdot$&$\cdot$&\SI{4.0e-14}{}&$\cdot$&\SI{1.0e-7}{}& \\
67&[58]&RFNVD&\SI{3.0e-9}{}&$\cdot$&$\cdot$&$\cdot$&$\cdot$&$\cdot$&\SI{1.0e-7}{}& \\
68&[58]&RFNVD&\SI{5.0e-9}{}&$\cdot$&$\cdot$&$\cdot$&$\cdot$&$\cdot$&\SI{1.1e-7}{}& \\
69&[62]&WGM&\SI{4.0e-5}{}&\SI{4.0e-5}{}&\SI{4.0e-5}{}&\SI{6.5e-14}{}&$\cdot$&$\cdot$&\SI{1.4e-7}{}& \\
70&[63]&MTJ&\SI{7.0e-6}{}&\SI{7.0e-6}{}&$\cdot$&$\cdot$&\SI{4.9e-11}{}&$\cdot$&\SI{1.5e-7}{}& \\
71&[64]&RFNVD&\SI{1.3e-7}{}&\SI{1.5e-7}{}&\SI{1.3e-7}{}&\SI{8.1e-21}{}&$\cdot$&$\cdot$&\SI{2.2e-7}{}& \\
72&[65]&RFNVD&\SI{5.0e-8}{}&\SI{5.0e-8}{}&\SI{5.0e-8}{}&\SI{1.3e-22}{}&$\cdot$&$\cdot$&\SI{2.9e-7}{}& \\
73&[66]&HALL&\SI{1.0e-6}{}&\SI{1.0e-6}{}&$\cdot$&$\cdot$&\SI{1.0e-12}{}&$\cdot$&\SI{3.0e-7}{}& \\
74&[67]&HALL&\SI{1.5e-6}{}&\SI{1.5e-6}{}&$\cdot$&$\cdot$&\SI{2.3e-12}{}&$\cdot$&\SI{6.0e-7}{}& \\
75&[59]&NVD&\SI{2.5e-8}{}&$\cdot$&$\cdot$&$\cdot$&$\cdot$&$\cdot$&\SI{6.0e-6}{}& \\
76&[54]&MFM&\SI{1.0e-8}{}&\SI{1.0e-8}{}&$\cdot$&$\cdot$&\SI{1.0e-16}{}&\SI{7.0e-20}{}&\SI{7.0e-4}{}& 
\end{tabular}

\end{center}
\caption{Continuation of Table~I.}
\end{table}

\clearpage

\nociteSI{SchmelzIEEETAS2016s} \nociteSI{DangAPL2010s} \nociteSI{StormAPL2017s} \nociteSI{SchmelzSST2011s} \nociteSI{KominisN2003s} \nociteSI{ShengPRL2013s} \nociteSI{DrungJAP1995s} \nociteSI{FaleySST2006s} \nociteSI{FaleyPP2012s} \nociteSI{Faley2013s} \nociteSI{GriffithOE2010s} \nociteSI{PannetierScience2004s} \nociteSI{BrandtPRL2000s} \nociteSI{LuomahaaraNC2014s} \nociteSI{KreyJAP1999s} \nociteSI{FongRSI2005s} \nociteSI{AwschalomAPL1988s} \nociteSI{RobbesSAA2006s} \nociteSI{VetoshkoSAAP2003s} \nociteSI{Vengalattore2007s} \nociteSI{CrepazSR2015s} \nociteSI{WolfPRX2015s} \nociteSI{FescenkoARX2019s} \nociteSI{KawaiTAS2016s} \nociteSI{Cromar1981s} \nociteSI{OdaEPS2016s} \nociteSI{VanHarlingenAPL1982s} \nociteSI{WoodPRA2015s} \nociteSI{SchwindtAPL2007s} \nociteSI{Sewell2012s} \nociteSI{Wang2012s} \nociteSI{SchmelzSST2017s} \nociteSI{JaspersePRA2017s} \nociteSI{EtoPRA2013s} \nociteSI{BarryPNAS2016s} \nociteSI{KimSR2016s} \nociteSI{Marauska2013s} \nociteSI{SchwindtAPL2004s} \nociteSI{DietscheNPhys2019s} \nociteSI{Ockeloen2013s} \nociteSI{GallopPCS2002s} \nociteSI{JengIEEEToM2012s} \nociteSI{WildermuthPRA2004s} \nociteSI{WildermuthN2005s} \nociteSI{ WildermuthAPL2006s} \nociteSI{ClevensonNP2015s} \nociteSI{HankardGGG2009s} \nociteSI{Behbood2013s} \nociteSI{GiazottoNP2010s} \nociteSI{WakaiAPL1988s} \nociteSI{Kirtley2016s} \nociteSI{Muessel2014s} \nociteSI{AhmadiPRA2017s} \nociteSI{KirtleyRPP2010s} \nociteSI{VasyukovNN2013s} \nociteSI{YangPRAp2017s} \nociteSI{FangPRL2013s} \nociteSI{LovchinskyS2016s} \nociteSI{MaletinskyNN2012s} \nociteSI{HuangAPL2014s} \nociteSI{BendingAP1999s} \nociteSI{ForstnerAM2014s} \nociteSI{LimaMST2014s} \nociteSI{ZhouARX2019s} \nociteSI{TrusheimNL2014s} \nociteSI{ChenaudJAP2016s} \nociteSI{OralITM2002s} 

\noindent
\section{References for Table~I. }
\label{sec:References2}


\renewcommand{\bibsection}{\subsection*{biblio name}}
\renewcommand{\bibsection}{}
\setcitestyle{numbers,square}

\bibliographystyleSI{./apsrev4-1no-url}
\bibliographySI{./MegaBib}

   

\end{widetext}

\end{document}